\documentclass[showpacs,twocolumn,prd]{revtex4-1}
\usepackage{slashed}
\usepackage{mathrsfs}
\usepackage{amsfonts}
\usepackage{amsmath}
\usepackage{amssymb}
\usepackage{revsymb}
\usepackage{graphicx,accents}
\usepackage{mathrsfs}
\usepackage{bm}
\usepackage{psfrag}
\usepackage{color}
\usepackage{hyperref}
\usepackage{natbib}
\usepackage{bbm}
\usepackage{bbold}

\usepackage{tabularx}

\usepackage[usenames,dvipsnames]{xcolor}

\newcommand{\be}{\begin{equation}}
\newcommand{\ee}{\end{equation}}
\newcommand{\ba}{\begin{array}}
\newcommand{\ea}{\end{array}}
\newcommand{\bea}{\begin{eqnarray}} 
\newcommand{\eea}{\end{eqnarray}} 
\newcommand{\bd}{\begin{displaymath}}
\newcommand{\ed}{\end{displaymath}}
\newcommand{\eps}{\varepsilon}

\newcommand{\mbf}[1]{\mathbf{#1}}
\newcommand{\trm}[1]{\textrm{#1}}

\newcommand{\figref}[1]{Fig. \ref{#1}}
\newcommand{\figrefa}[1]{Fig. \ref{#1}a}
\newcommand{\figrefb}[1]{Fig. \ref{#1}b}

\newcommand{\eqnref}[1]{Eq. (\ref{#1})}

\newcommand{\sxnref}[1]{Sec. \ref{#1}}
\newcommand{\appref}[1]{Appendix \ref{#1}}

\newcommand{\tsf}[1]{\textsf{#1}}

\newcommand{\vkap}{\varkappa}
\newcommand{\vphi}{\varphi}

\newcommand{\Ai}{\trm{Ai}}

\newcommand{\ef}{\mathcal{E}}

\newcommand{\nn}{\nonumber}

\newcommand{\e}{\mathbb{e}}

\makeatletter
\DeclareRobustCommand{\cev}[1]{%
  \mathpalette\do@cev{#1}%
}
\newcommand{\do@cev}[2]{%
  \fix@cev{#1}{+}%
  \reflectbox{$\m@th#1\vec{\reflectbox{$\fix@cev{#1}{-}\m@th#1#2\fix@cev{#1}{+}$}}$}%
  \fix@cev{#1}{-}%
}
\newcommand{\fix@cev}[2]{%
  \ifx#1\displaystyle
    \mkern#23mu
  \else
    \ifx#1\textstyle
      \mkern#23mu
    \else
      \ifx#1\scriptstyle
        \mkern#22mu
      \else
        \mkern#22mu
      \fi
    \fi
  \fi
}

\newcommand*\xbar[1]{%
  \hbox{%
    \vbox{%
      \hrule height 0.5pt 
      \kern0.2ex
      \hbox{%
        \kern-0.15em
        \ensuremath{#1}%
        \kern-0.15em
      }%
    }%
  }%
}

\newcommand{\optional}[1]{}

\graphicspath{ {./} }

\begin{document}
\title{A uniform locally constant field approximation \\ for photon-seeded pair production}
 
\author{B.~King}
\affiliation{Centre for Mathematical Sciences, University of Plymouth, Plymouth, PL4 8AA, United 
Kingdom}
\email{b.king@plymouth.ac.uk}


\date{\today}
\begin{abstract}
A challenge to upcoming experiments that plan to collide a particle beam with laser pulses of moderate intensity is how to correctly incorporate quantum effects into simulation frameworks. Using a uniform approach, we extend the widely-used locally constant field approximation (LCFA) to derive an improved rate of photon-seeded pair creation (the nonlinear Breit-Wheeler process). By benchmarking our ``ULCFA'' expressions with the lightfront spectrum of: i) exact analytical results and ii) numerical integration of the QED probability for short pulses, we show that our extended approach remains accurate at smaller values of the intensity parameter than the standard LCFA. 
\end{abstract}
\maketitle
%
%
%
%
%
%
%
%
The presence of an electromagnetic (EM) background allows for processes that are otherwise kinematically forbidden. The example we study here is the decay of a photon into an electron-positron pair in a plane-wave background. If the background is sufficiently weak, the photon decay can be understood as a $\gamma$-$\gamma$ collision where one photon originates from the backgound. If the centre-of-mass energy is above the pair creation threshold $2mc^{2}$, the photon can decay to an electron-positron pair via the linear process first studied by Breit and Wheeler \cite{breit34}. However, if the background is intense enough that many field quanta participate in the decay of the seed photon, the background can be modelled as classical and interactions between the photon and background must in general be included to all orders in the intensity parameter squared $\xi^2$ (proportional to the fine-structure constant $\alpha\approx 1/137$), which describes the \emph{nonlinear} Breit-Wheeler process. Photon-seeded pair creation has been studied in EM backgrounds that are constant crossed \cite{nikishov64, narozhny69}, monochromatic \cite{nikishov64, narozhny69} and purely electric \cite{brezin70}. More recently, the effect of a finite pulse duration has been explored \cite{heinzl10,nousch12,titov12,titov19} and the higher frequency modes found to be beneficial for pair-creation. The effect of laser pulse focussing for high-energy photons has also been recently calculated \cite{dipiazza16}. (Reviews of high-intensity QED in lasers can be found in \cite{dipiazza12,king15a,narozhny15,Hu:2019dij}).
\newline

One reason the locally constant field approximation (LCFA) is so called, is because of its equivalence to taking the instantaneous lightfront rate for a single (dressed) vertex process in a constant crossed field (CCF), inserting the dependency on a given background field, and then integrating the rate over the phase-dependency of that background. This has been shown to be equivalent to a Taylor expansion in the interference phase variable \cite{harvey15,dipiazza18a,king19a} and hence can capture average spectral structure, but not harmonic substructure \cite{ritus85,harvey15}. In laser-based high-intensity QED, the LCFA has been most studied in its application to nonlinear Compton scattering \cite{nikishov64, harvey15}, where a scheme has been developed to patch it at low lightfront momentum to the perturbative Klein-Nishina result \cite{dipiazza18a}, and has been extended with higher-order derivatives in the form of the ``LCFA+'', to be accurate at lower field intensity values \cite{king19a}. The angular LCFA (ALCFA), which was first formulated some time ago \cite{baier94}, has recently been investigated in numerical simulations \cite{blackburn19, dipiazza19}. The LCFA has also been applied to the process of photon-seeded pair-creation \cite{dipiazza19}, to spontaneous pair-production from vacuum \cite{aleksandrov19} and most recently, to single photon absorption \cite{king19b} and pair-annihilation \cite{king19c} in plane waves. In addition to the theoretical investigations, the LCFA underpins particle-in-cell Monte Carlo simulation \cite{nerush11,elkina11,ridgers12,king13a,bulanov13,ridgers14,blackburn15,gelfer15,jirka16,gonoskov17,efimenko19} of quantum effects in high-intensity laser-plasma experiments \cite{sarri14,sarri15,cole18,poder18}. Furthermore, the LCFA is also applied in other fields, such as in astrophysics \cite{harding06} and beamstrahlung \cite{yokoya92}.
\newline

In the current paper we use a uniform Airy approximation \cite{miller53, soares10, Dunne:2014bca} to extend the LCFA by including higher order field derivatives at the level of the functional arguments of the LCFA. We will refer to this as the ``ULCFA'', despite its similarity to the ``LCFA+'' for nonlinear Compton scattering introduced in \cite{king19a}, because the ULCFA does not involve an expansion of exponential terms, and so the higher-order derivatives are included to all orders in the LCFA arguments. We then apply our ULCFA to photon-seeded pair-creation and by benchmarking it  against the exact analytical result for pair-creation in a circularly-polarised monochromatic background, as well as numerical evaluation of QED expressions for a finite short $\cos^{2}$ pulse, demonstrate the ULCFA to be consistently more accurate than the LCFA at lower values of the laser intensity. For the short-pulse case, we identify the position of the harmonic resonances that provide the sub-structure in the lightfront spectrum, which we show can be found by solving the kinematics for the cycle-averaged intensity of the background. Of particular focus throughout is the transition region as the laser pulse intensity is increased from perturbative, multiphoton pair-creation to that of ``all-order'' non-perturbative pair-creation at small coupling. Motivation for this focus originates in upcoming experiments such as LUXE at DESY \cite{altarelli19} and E320 at FACET-II \cite{slacref1}, planning to probe high-intensity QED effects by colliding high-energy electron beams with intense optical laser pulses, thereby extending the results of the seminal E144 experiment at SLAC \cite{bula96,burke97,bamber99} more than two decades ago. These experiments are complementary to those using laser-wakefield-accelerated electron pulses, such as future experiments at next-generation high-intensity laser facilities such as ELI \cite{eli16} and the Station of Extreme Light (SEL) \cite{shen18} at the Shanghai Superintense Ultrafast Laser Facility (SULF).
\newline

The paper is organised as follows. In Sec. I we outline the derivation of the ULCFA for pair-creation using a uniform Airy approach.  In Sec. II we compare the ULCFA and the LCFA with exact solutions at the level of the lightfront spectrum, for i) a circularly-polarised monochromatic background and ii) the numerical integration of the full QED probabilty for a short $\cos^{2}$ pulse. In Sec. III we conclude. Appendix A provides detail on the numerical method for integration of the short-pulse QED probability.

\section{Uniform LCFA}
The scattering matrix element for photon-seeded pair-creation in a plane-wave background is:
\bea
\tsf{S}_{\tsf{fi}} = i~\int d^{4}x~ \overline{\psi}_{p}\slashed{a}_{\gamma} \psi_{q}^{+},
\eea
where we recall the Volkov solutions to the Dirac equation for a plane-wave EM background:
\bea
\overline{\psi}_{p} &=& \overline{E}_{p}(\vphi)\,\e^{ip\cdot x + iS_{p}(\vphi)}\nn\\
\psi_{q}^{+} &=& E_{-q}(\vphi)\,\e^{iq\cdot x + iS_{-q}(\vphi)}\nn\\
E_{p}(\vphi) &=& \left(1+ \frac{\slashed{\vkap}\slashed{a}(\vphi)}{2\,\vkap\cdot p}\right)\,\frac{u_{r}(p)}{\sqrt{2p^{0}V}} \nn \\
S_{p}(\vphi) &=& \int^{\vphi} \frac{2 p\cdot a(\psi) - a^{2}(\psi)}{2\,\vkap\cdot p}~d\psi,\nn
\eea
using the lightlike external-field wavevector $\vkap$, phase $\vphi = \vkap \cdot x$, and the rescaled potential $a^{\mu} = eA^{\mu}$ with positron charge $e>0$, where $\bar{u}_{r}(p)u_{s}(p) = 2 \delta_{r,s}$, $\sum_{r}\bar{u}_{r}(p)u_{r}(p)=\slashed{p}+m$ and $p$ ($q$) is the electron (positron) four-momentum. We employ an incident plane-wave photon state with momentum $k$:
\bea
a_{\gamma} = \frac{e \eps}{\sqrt{2k^{0}V}}\mbox{e}^{ik\cdot x},
\eea
and photon polarisation $\eps\cdot\eps = -1$.
\newline

For a plane wave with scaled vector potential $a = eA$, the probability $\tsf{P}$, for pair-creation given by
\[
\tsf{P} = V^{2}\int \frac{d^{3}p\, d^{3}q }{(2\pi)^{6}}~\sum_{\tsf{spin}}\langle|\tsf{S}_{\tsf{fi}}|^{2}\rangle_{\tsf{pol}},
\]
where a sum is performed over outgoing fermion spins and an average over incoming photon polarisation, is found to be:
\bea
\tsf{P} &=& \lim_{\eps\to0}\frac{\alpha}{\eta_{k}}\frac{i}{2\pi}\int_{-\infty}^{\infty} d\phi \int_{-\infty}^{\infty} d\theta\int_{0}^{1}dt~\frac{\mbox{e}^{\frac{i\theta\mu(\theta)}{2\eta_{k}t(1-t)}}}{\theta+i\eps} \label{eqn:Ppulse}\\
&& \left\{1+\frac{\left[a(\sigma + \frac{\theta}{2})-a(\sigma - \frac{\theta}{2})\right]^{2}}{4m^{2}}\left(-2 + \frac{1}{t(1-t)}\right)\right\}\nn
\eea
where the lighfront momentum fraction $t=\eta_{p}/\eta_{k}$, $\eta_{p} = \vkap\cdot p/m^{2}$ etc. $\vphi = \vkap\cdot x$, the normalised Kibble mass is given by:
\[
\mu(\theta) = 1 - \frac{1}{\theta}\int^{\sigma+\frac{\theta}{2}}_{\sigma-\frac{\theta}{2}} \frac{a^{2}(y)}{m^{2}} dy + \left[\frac{1}{\theta}\int^{\sigma+\frac{\theta}{2}}_{\sigma-\frac{\theta}{2}} \frac{a(y)}{m} dy\right]^{2}
\]
and $\theta = \vphi - \vphi'$ and $\phi = (\vphi+\vphi')/2$ where $\vphi$ and $\vphi'$ are the original external-field phase variables. To derive the standard LCFA from \eqnref{eqn:Ppulse}, one i) performs a Taylor expansion of the Kibble mass in $\theta$, retaining the lowest-order terms that ensure convergence of the integral, namely to order $\theta^{2}$; ii) expands the pre-exponent to order $\theta^{2}$, i.e. making the replacements:
\bea
\theta\mu &\to& \theta + \frac{\ef^{2}}{12}\theta^{3}; \quad \left(a(\phi)-a(\phi')\right)^{2} \to -\theta^{2}\ef^{2},  \label{eqn:LCFArep1}
\eea
where the dimensionless field strength $\pmb{\ef}(\phi)$ is defined through $a'(\phi) =  m(0, \pmb{\ef}(\phi))$, 
and then iii) performs the remaining $\theta$-integral, to arrive at:
\bea
\tsf{P}^{\tsf{LCFA}}_{e} = \frac{\alpha}{\eta_{k}} \int_{-\infty}^{\infty} d\phi\,\int_{0}^{1} dt ~\mathcal{I}^{\scriptsize \textsf{LCFA}} \nn
\eea
\bea
\mathcal{I}^{\scriptsize \textsf{LCFA}} &=& \Ai_{1}(z_{e})+\left(\frac{2}{z_{e}} - \xi\eta_{k}\sqrt{z_{e}}\right)\Ai'(z_{e}) , \label{eqn:lcfa}
\eea
where we define:
\bea
z_{e} = \left(\frac{1}{|\pmb{\ef}(\phi)| \eta_{k} t(1-t)}\right)^{2/3},
\eea
which agrees with the CCF result as expected \cite{nikishov64}. We quantify the magnitude of the field strength with the parameter $\xi(\phi)\geq0 $ via the definition $|\pmb{\ef}(\phi)| = m\xi(\phi)$. 
\newline

To go beyond the LCFA, we study an integral with the typical exponential form of a QED pulse expression \eqnref{eqn:Ppulse}:
\bea
\mathcal{J} = \int_{-\infty}^{\infty}d \theta ~  \e^{i\lambda g(\theta)},
\eea
where $g$ is a real function and $\lambda\gg 1$ is an asymptotic parameter.  Adapting the discussion in \cite{soares10}, we can cast the integrand in the form of an Airy kernel by choosing
\bea
\tilde{g}(y) = g[\theta(y)] =  X^{2} y + \frac{y^{3}}{3},
\eea
where
\bea
X^{2} = \frac{1}{\lambda}\left\{\frac{3}{4}\left[g(\theta_{i,1}^{\ast}) -g(\theta_{i,2}^{\ast})\right]\right\}^{2/3}, \label{eqn:airyarg1}
\eea
for stationary points $\theta_{i,1}^{\ast} = - \theta_{i,2}^{\ast}$. Then, considering the stationary points pairwise in this way, we can write:
\bea
\mathcal{J} \sim \sum_{i} \left(\frac{d\theta(y_{i,1}^{\ast})}{dy}+\frac{d\theta(y_{i,2}^{\ast})}{dy}\right) \int_{-\infty}^{\infty} \e^{i\lambda(X^{2}y + \frac{y^{3}}{3})}\, dy,\label{eqn:jexp1}\nn \\
\eea
where the stationary points in the new variable $y$ fulfill $\theta(y^{\ast}_{i,j}) = \theta^{\ast}_{i,j}$ and the sum is over all pairs of stationary points.
\newline

To extend the LCFA then, we apply the previous general discussion to the same form of exponent as in the pulse probability \eqnref{eqn:Ppulse}. Setting the dummy variable $\lambda=1$ and performing a Taylor expansion of the exponent in $\theta$, but now up to $\theta^{5}$, we find:
\bea
g(\theta) = \frac{1}{2\eta_{k}t(1-t)}\left[\theta + \frac{1}{12}f_{0} \theta^{3} + \frac{1}{720}f_{1}\theta^{5}\right], \label{eqn:T1}
\eea
with $f_{0} = \pmb{\ef}\cdot \pmb{\ef}$ and $f_{1} =  \pmb{\ef}'^{2} + 3\,\pmb{\ef}\cdot  \pmb{\ef}''$. This leads to two pairs of stationary points. One of these pairs leads to a large Airy argument in the region of interest, and therefore a suppressed contribution, which we discard. 
Applying \eqnref{eqn:airyarg1} to the remaining pair of stationary points, equates to the prescription of replacing Airy-function arguments via $z_{e} \to z_{e}^{\tsf{+}}$, where:
\bea
z_{e}^{\tsf{+}} = z_{e}\,\left(1+ \theta\left[\xi(\phi)-\xi^{\ast}\right]\frac{\pmb{\ef}'^{2}(\phi) + 3\pmb{\ef}(\phi)\cdot \pmb{\ef}''(\phi)}{30|\pmb{\ef}(\phi)|^{4}}\right)^{2/3}. \nn \\ \label{eqn:zplus}
\eea
Here, we have manually added a filter to the Airy argument (as opposed to outside the Airy function \cite{king19a}), which ensures the local $\xi$ is greater than some positive $\xi^{\ast}$. The significance and necessity of the filter will become clearer in \sxnref{sxn:circ}, but it can be understood at this point by the fact that the Taylor expansion in \eqnref{eqn:T1} requires $\xi \not \ll 1$ \cite{king19a}). Such a filter appears to be a standard consequence of including derivative corrections to the LCFA, whether at the level of the intensity \cite{king19a}, or in particle energy \cite{dipiazza19} (where higher derivatives can be used to define a new timescale to use in sampling the ``local field'').

To deal with the pre-exponent in \eqnref{eqn:jexp1}, through repeated use of l'H\^opital's rule, we find:
\[
 \frac{d\theta(y_{j}^{\ast})}{dy} = \left(\frac{\tilde{g}'''(y_{j}^{\ast})}{g'''(\theta_{j}^{\ast})}\right)^{1/3} =  \frac{2}{\xi \sqrt{\widehat{z}_{e}}}\]
\bea
\widehat{z}_{e} = z_{e}\left[1 + \frac{8}{3 } \frac{( \pmb{\ef}'\cdot \pmb{\ef}' + 3\,\pmb{\ef}\cdot  \pmb{\ef}'')}{| \pmb{\ef}\cdot \pmb{\ef}|^{3}}\right]^{2/3}, \label{eqn:preexp1}
\eea
and since $d\theta(y_{i,1}^{\ast})/dy=d\theta(y_{i,2}^{\ast})/dy$, we acquire the same prefactor as in the derivation of the LCFA, modified by higher derivative terms. At this point, we ignore pre-exponent corrections to the LCFA, because we are assuming $\xi \gg 1$ and as we can see by expanding \eqnref{eqn:preexp1} in powers of $1/\xi$, the pre-exponent corrections scale as $1/\xi^{4}$, whereas the Airy-argument corrections scale as $1/\xi^{2}$ and hence are more significant. 
\newline

Then we define the ULCFA integrand for electron-seeded pair-creation as:
\bea
\mathcal{I}^{\scriptsize \textsf{ULCFA}} &=& \Ai_{1}(z^{\tsf{+}}_{e})+\left(\frac{2}{z_{e}} - \xi\eta_{k}\sqrt{z_{e}}\right)\Ai'(z^{\tsf{+}}_{e}) . \nn \\
\eea
(cf. \eqnref{eqn:lcfa}). We note that the higher derivatives of the external field can be written in terms of the local dynamics of the electron using the Frenet-Serret formalism \cite{Seipt:2019dnn}, where the instantaneous four-acceleration of the electron in the plane-wave background, $\dot{u}(\phi)$, is related to the combination $\dot{u}(\phi) = \eta_{p}\xi(\phi)$, and higher derivatives of the four-velocity $u$ can be used to write, e.g.
\[
 \frac{\pmb{\ef}'^{2}(\phi) + 3\pmb{\ef}(\phi)\cdot \pmb{\ef}''(\phi)}{30|\pmb{\ef}(\phi)|^{4}} = - \frac{\ddot{u}\cdot\ddot{u} + 3\, \dot{u}\cdot\dddot{u}}{30\,\dot{u}^{4}},
\]
without explicit reference to the field.
\newline

In the following, we evaluate the effectiveness of the ULCFA by direct calculation and comparison with exact and numerical solutions.

\section{Benchmarking to exact solutions}
The probability for $n$ background photons to contribute to pair-creation is proportional to $\xi^{2n}$. Therefore when $\xi \sim O(1)$, we expect the harmonic structure in the pair-spectrum to be most pronounced. (If $\xi\gg1$, many harmonics of narrow harmonic range contribute and merge into a continuum, whereas if $\xi \ll 1$, only the leading-order harmonic contributes.) This harmonic structure is absent in a CCF, and therefore also absent from the LCFA. So to test the applicability of the LCFA and ULCFA, we compute the lightfront spectrum in this regime. We first benchmark the ``instantaneous'' expression, $\mathcal{I}$, defined above, to be the non-trivial part of the ``instantaneous probability'', where in general:
\[
 \mathcal{I} = \mathcal{I}(\phi, t, \xi, \eta),
\]
against an analytical solution, before comparing with a numerical evaluation of the full pulsed result \eqnref{eqn:Ppulse}.

\subsection{Circularly-polarised monochromatic background} \label{sxn:circ}
We define a circularly-polarised monochromatic background through the scaled vector potential:
\bea
 a = (0,\mbf{a}^{\perp}, 0); \quad \mbf{a}^{\perp} = m\xi \left[\cos\phi, \sin\phi\right]. \label{eqn:acirc}
\eea
Applying this to the ULCFA argument \eqnref{eqn:zplus}, it follows that
\[
 z_{e}^{\tsf{+}} = z_{e}\left(1 - \frac{1}{15\xi^{2}}\right)^{2/3}.
\]
(The filter on $\xi(\phi)$ in \eqnref{eqn:zplus} has been neglected, because in a circularly-polarised monochromatic background, $\xi$ is a constant.) Interestingly, if we then calculate the asymptotic limit for small quantum nonlinearity parameter, $\chi_{k}=\xi\eta_{k} \to 0$, we find:
\bea
 \mathcal{I}^{\scriptsize \textsf{ULCFA}} \sim \chi_{k}\exp\left[-\frac{8}{3\chi_{k}}\left(1-\frac{1}{15\xi^{2}}\right)\right],
\eea
which is the CCF limit of the monochromatic expression for pair-creation in the limit of large harmonic order \footnote{For example, on P. 551 of the review \cite{ritus85}, also reported in \cite{hartin19}.}. For $\chi_{k}\ll1$, it is known \cite{ritus85} to be an accurate approximation when $\xi^{3}\eta_{k} \gg 1$. So whilst we expect the ULCFA to be useful when $\xi > 1$, it can also be accurate when $\xi <1$, provided the energy of the seed photon is sufficiently high (we note $\eta_{k}\sim1$ for a typical optical laser wavelength $800\,\trm{nm}$, when the seed photon has energy of the order of $100\,\trm{GeV}$). 
\newline

Another illustration of the analytical limit where we expect the ULCFA to be accurate comes from considering what happens if one specifies the pulsed result \eqnref{eqn:Ppulse} to a linearly-polarised monochromatic background, which has the form $\mbf{a}^{\perp} = m\xi[\sin\phi,0]$, instead. By performing an asymptotic analysis for $\chi_{k} \to 0$ in the spirit of \cite{torgrimsson18, torgrimsson19}, then for the saddle at $\phi=0$, $t=1/2$, $\theta = 2i\xi\cosh^{-1}[\sqrt{1+\xi^{2}}/\xi]$, the exponent becomes:
\bea
 \mathcal{I} \sim \exp\left\{\frac{2\xi}{\chi_{k}}\left(\sqrt{1+\xi^{2}} - (2+\xi^{2})\,\trm{sech}^{-1}\frac{\xi}{\sqrt{1+\xi^{2}}}\right) \right\}. \nn \\ \label{eqn:monosaddle1}
\eea
If one also considers strong field strengths by taking the limit of large $\xi$ in the small $\chi_{k}$ limit \footnote{Recent work has shown that the order in which such limits are taken is crucial, and that they are not, in general, commutative \cite{ilderton19a}.}, this becomes:
\[
 \mathcal{I} \sim \exp\left\{-\frac{8}{3\chi_{k}}\left(1-\frac{1}{10\xi^{2}}\right) \right\},
\]
which is also exactly what one acquires specifying the ULCFA argument \eqnref{eqn:zplus} to this background, and replacing the $\phi$-dependent argument with the saddle point at $\phi=0$. From the saddle-point analysis, we note we also have access to small $\xi$ in the small $\chi_{k}$ limit, in which case \eqnref{eqn:monosaddle1} reduces to:
\bea
 \mathcal{I} \sim \xi^{2n^{\ast}}; \quad 
n^{\ast} =  \frac{2(1+\xi^{2})}{\eta_{k}}, \label{eqn:nast1}
\eea
which is exactly the leading-order multiphoton term for pair-creation (this will be immediately relevant when studying the circularly-polarised monochromatic rate below, where $n^{\ast}$ will be replaced with the nearest greater integer $\lceil n^{\ast} \rceil$). Therefore, \eqnref{eqn:monosaddle1} can interpolate between the tunneling and multiphoton regime, for small $\chi_{k}$. This is an analytic example of the interpolating function used in the analysis of the SLAC E144 experiment \cite{bamber99}, and is similar to the seminal result by Brezin and Itzykson for pair-creation by an alternating field \cite{brezin70}.
\newline

We write the probability for pair-creation in a circularly-polarised background \cite{nikishov64} in the same form as for the LCFA:
\bea
\tsf{P}^{\tsf{mono}}_{e} = \sum_{n\geq n^{\ast}}^{\infty} \tsf{P}^{\tsf{mono}}_{e,(n)}; \quad \tsf{P}^{\tsf{mono}}_{e,(n)} = \frac{\alpha}{\eta_{k}} \int_{-\infty}^{\infty} d\phi\,\int_{t_{-}}^{t_{+}} dt ~\mathcal{I}^{\scriptsize \textsf{mono}}_{n}\nn 
\eea
\bea
\mathcal{I}^{\scriptsize \textsf{mono}}_{n} &=&  J_{n}^{2}(z_{e,(n)})- \frac{\xi^{2}}{4}\left(2 - \frac{1}{t(1-t)} \right)\left[J_{n+1}^{2}(z_{e,(n)})\right. \nn \\ && \left. +J_{n-1}^{2}(z_{e,(n)})-2\,J_{n}^{2}(z_{e,(n)})\right],\nn
\eea
\bea
 z_{e,(n)}= \frac{\xi\sqrt{1+\xi^{2}}}{\eta t(1-t)}\sqrt{4t(1-t)\frac{n}{n^{\ast}}-1} \label{eqn:Pmono1}
\eea
where $J_{n}$ is the $n$th-order Bessel's function of the first kind \cite{olver97}, the threshold harmonic is given by  $\lceil n^{\ast} \rceil $  where $n^{\ast}$ is defined in \eqnref{eqn:nast1} and the harmonic range is given by:
\bea
t_{-} < t < t_{+}; \qquad t_{\pm} = \frac{1}{2}\left(1 \pm \sqrt{1-\frac{n^{\ast}}{n}}\right).
\eea
Although there is an integration over phase $\phi$ in the monochromatic probability \eqnref{eqn:Pmono1}, no part of the integrand depends on $\phi$. Therefore, it is customary for $\partial\tsf{P}_{e}^{\tsf{mono}}/\partial\phi$ to be interpreted as a probability rate per unit field cycle. 
\begin{figure}[h!!]
 \includegraphics[width=8cm]{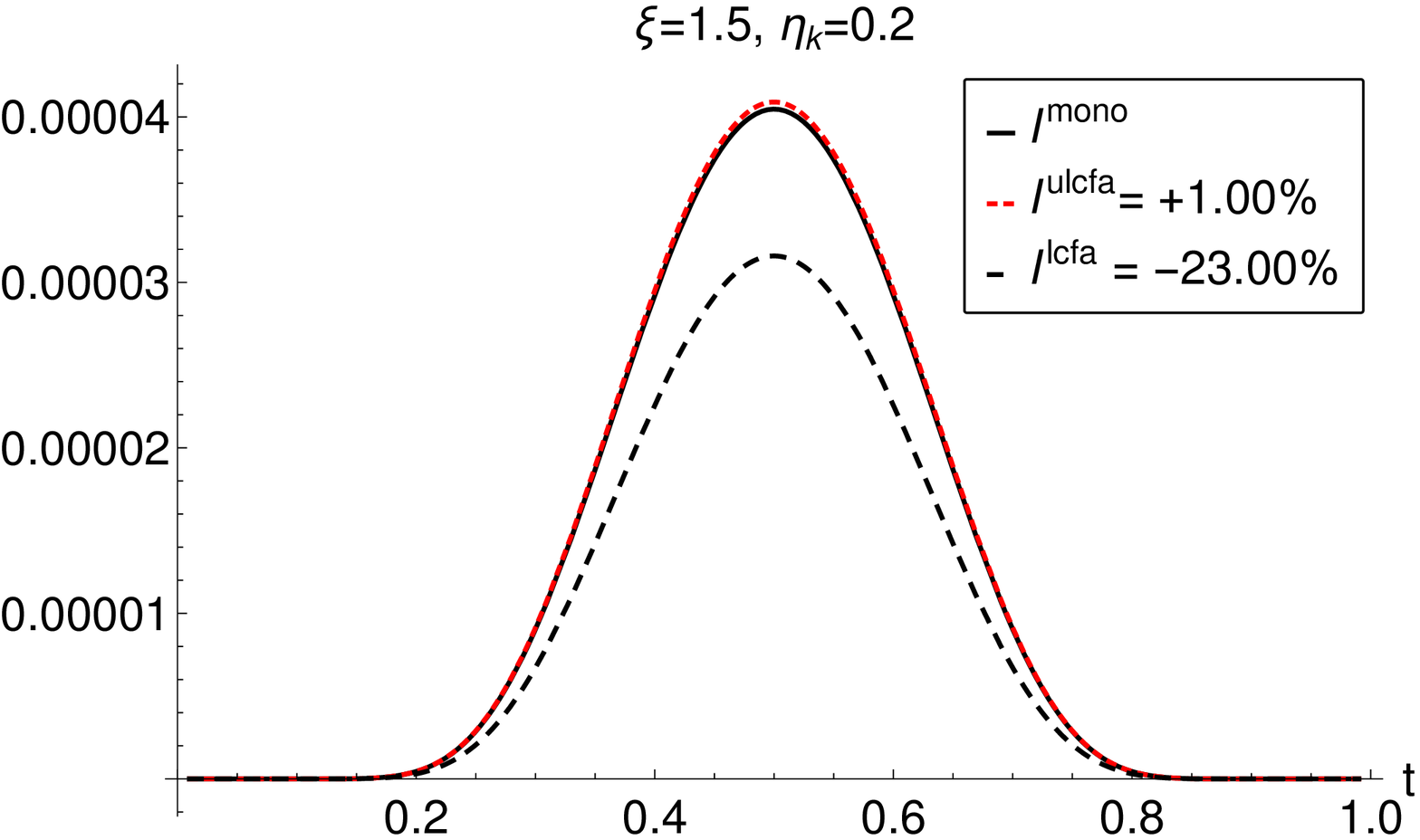}\\[3ex]
 \includegraphics[width=8cm]{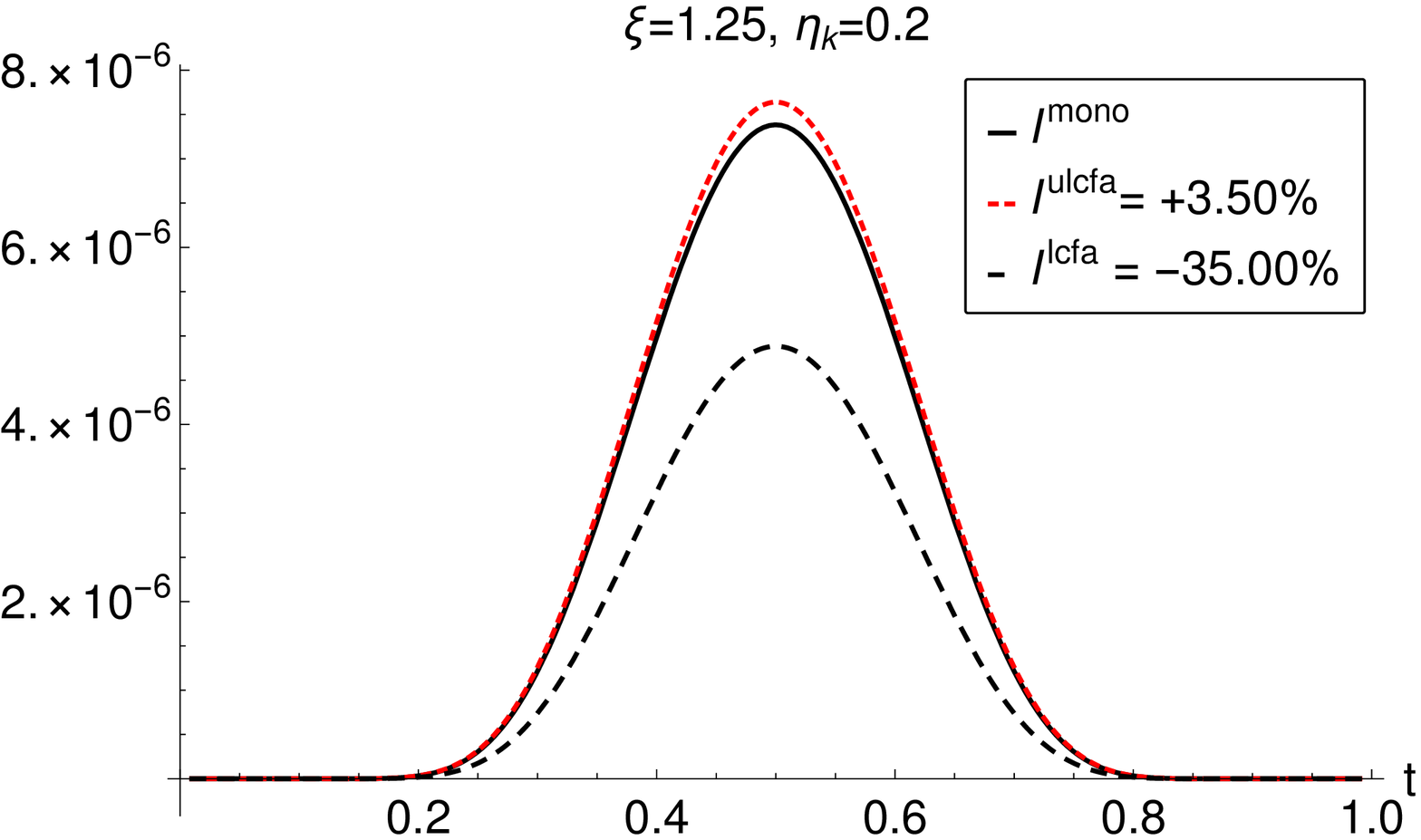}\\[3ex]
 \includegraphics[width=8cm]{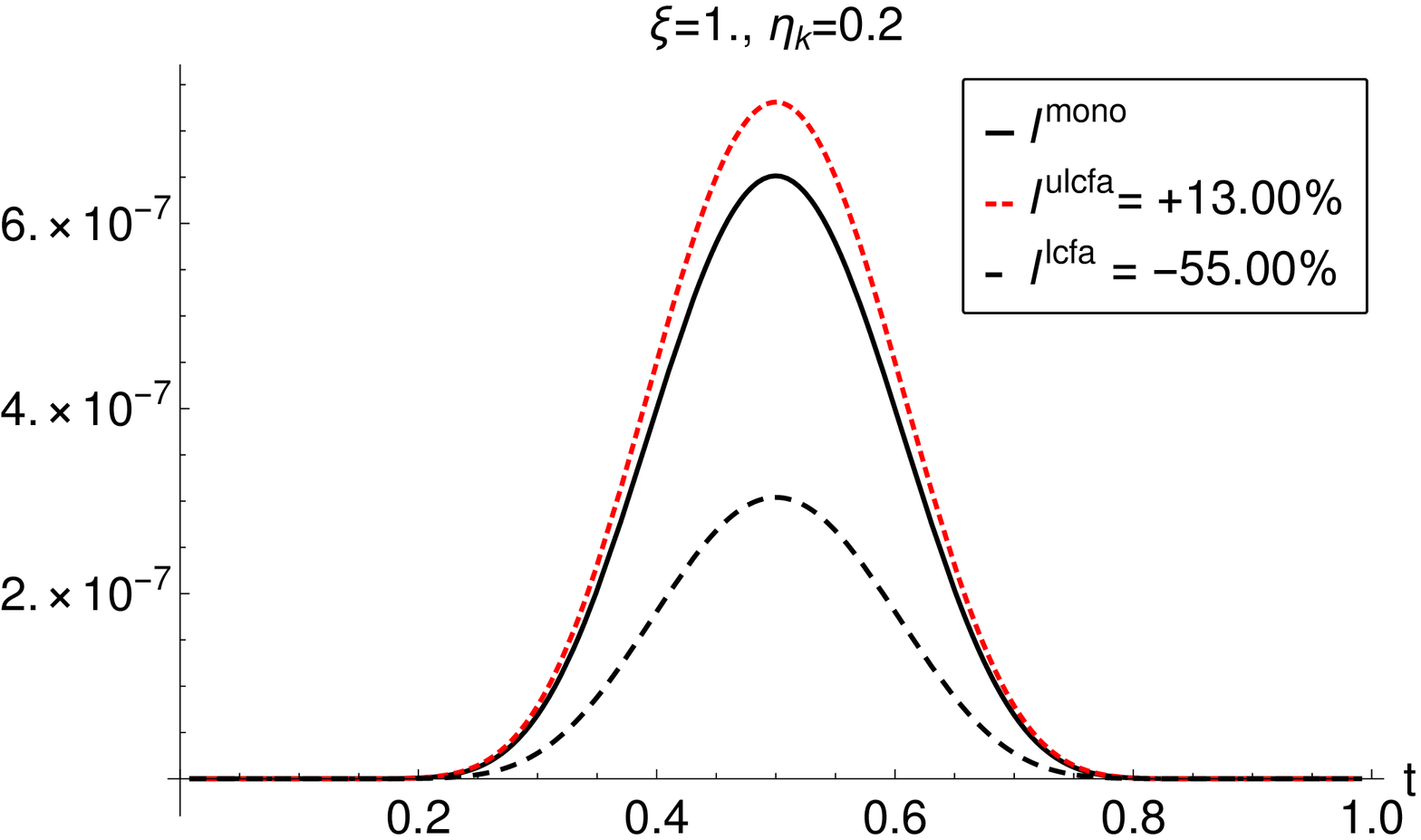}
 \caption{Comparison of the LCFA and ULCFA to the exact analytical result for a circularly-polarised monochromatic background. The percentages in boxes are the relative errors in the integral, calculated by integrating the curve over $t$. The relative error is further investigated in \figref{fig:mono2}.} \label{fig:mono1}
\end{figure}

\begin{figure}[h!!]
\includegraphics[width=6cm]{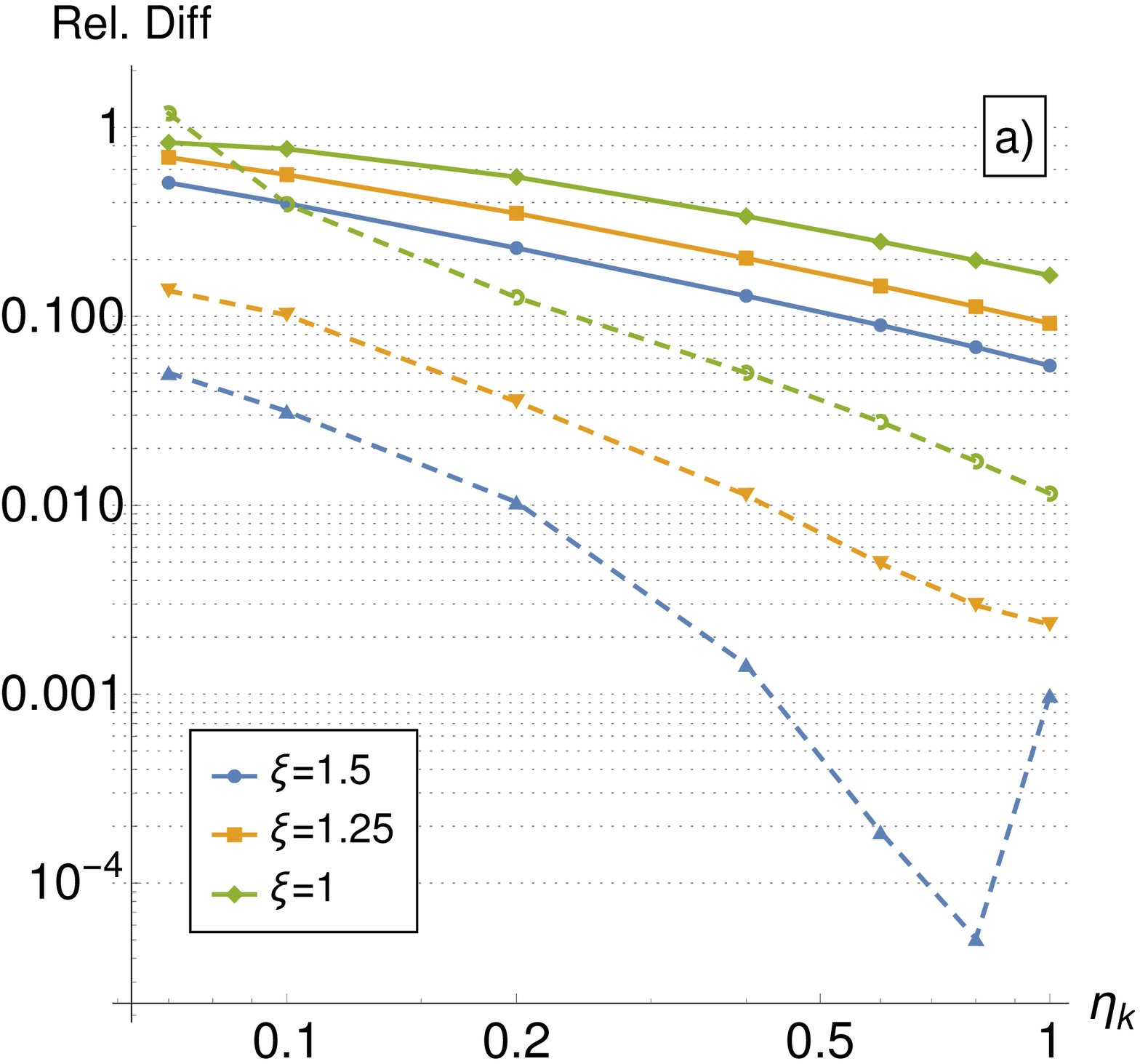} \\
\includegraphics[width=6cm]{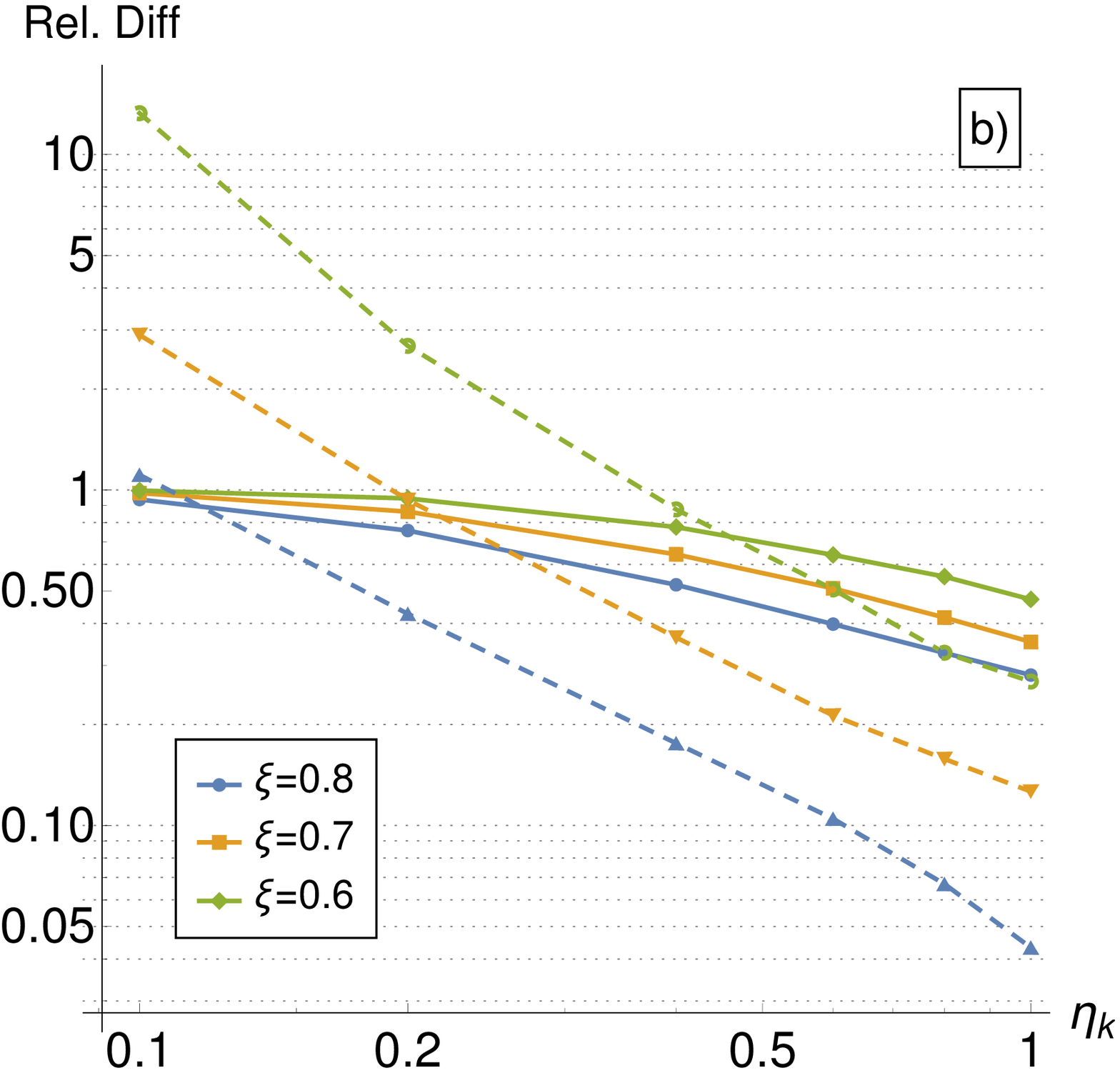} 
\caption{A comparison of the absolute relative error of the LCFA (solid lines) and ULCFA (dashed lines), compared to the analytical monochromatic result of the integrated rate $|\int\,\mathcal{I}\,dt/\int\,\mathcal{I}^{\tsf{mono}}\,dt-1|$.} \label{fig:mono2}
\end{figure}

Example spectra are compared in \figref{fig:mono1}. In general, the LCFA and ULCFA both increase in accuracy as $\xi$ is increased from $\xi=1$. However, it can be seen that the ULCFA is substantially more accurate than the LCFA in the region close to $\xi \approx 1$, where the LCFA is not expected to be accurate. In \figrefa{fig:mono2}, it is demonstrated how this accuracy also depends on the energy parameter $\eta_{k}$. If $\xi$ is reduced below $\xi=1$, the ULCFA begins to be less accurate than the LCFA. In \figrefb{fig:mono2}, it is shown that for small energy parameter, the ULCFA develops large errors. 
Here we see the motivation for introducing the intensity filter in the definition of the ULCFA in \eqnref{eqn:zplus}. If $\xi$ and $\eta$ are small, na\"ively using the modified Airy arguments can lead to a poor approximation.  However, this is easy to accomodate for, by using an intensity filter, such  that when $\xi(\phi) <\xi^{\ast}$, the LCFA is chosen in preference to the ULCFA. To determine $\xi^{\ast}$, one could use the intersection of the ULCFA and LCFA curves in \figrefb{fig:mono2} for whichever seed photon energy was being used. Because we are interested in demonstrating the efficacy of the ULCFA in modelling upcoming high-energy experiments such as LUXE and E320, where $\eta_{k} \sim O(0.1)$, in the following, we simply set $\xi^{\ast}=0.7$ without recourse to optimisation.

\subsection{Numerical evaluation of pulse integral}
In studies of the accuracy of the LCFA in modelling nonlinear Compton scattering, it was noted \cite{harvey15} that since the LCFA only samples ``local'' parts of the background, it cannot include significant interference effects, which are required to produce a harmonic structure. However, the situation in pair-creaton is different, because the mass gap defines a threshold harmonic, which, for electron beam energies available in upcoming particle-physics experiments (of the order of $10\,\trm{GeV}$), is already of the order of $n\sim O(10)$. We know that as $n \to \infty$, the Bessel functions in  monochromatic rates for nonlinear Compton and pair-creation can be well-approximated by their asymptotic limit \cite{seipt17,nist_dlmf}, which are Airy functions, corresponding to the CCF result. So we could expect that the LCFA and ULCFA can be potentially better approximations for pair-creation than nonlinear Compton scattering, which has no such limitation on the minimum contributing harmonic, other than $n>0$. However, there is an extra source of structure in finite pulsed backgrounds, which originates from the pulse envelope. It was shown \cite{heinzl10} that when the laser-averaged lightfront momentum transfer is a multiple of the laser carrier frequency, a finite-pulse harmonic resonance effect causes sub peaks to appear in the spectrum. This clearly cannot be modelled by the LCFA nor by monochromatic rates. 
\newline

In order to investigate how well the ULCFA performs for finite pulses, we use a finite $\cos^{2}$ pulse with scaled vector potential:
\bea
 a = (0,a^{\perp},0, 0); \quad a^{\perp} = m\xi \cos^{2}\left(\frac{\pi\phi}{2\Phi}\right)\,\sin\phi, \label{eqn:apulse2}
\eea
$\Phi = 2\pi N$, $N\in \mathbb{Z}$ and $|\phi|<\Phi$ (we have specifically chosen a potential that vanishes asymptotically to avoid infra-red effects \cite{dinu12}). For circularly-polarised backgrounds, we might expect the ULCFA to be particularly good because $a^{2}$ is a constant and the only $\phi$-dependent term in the Kibble mass is the term linear in $a$. For this reason, we test the ULCFA with the above \emph{linearly}-polarised pulse.  We then compare the phase-integrated spectrum, $\mathcal{S}$, defined e.g. for the ULCFA as:
\bea
\mathcal{S}^{\scriptsize \textsf{ULCFA}} &=& \int_{-\infty}^{\infty} d\phi ~ \mathcal{I}^{\tiny\tsf{ULCFA}}(\phi),
\eea
where we note that now, $\mathcal{S}$ is a function of:
\[
 \mathcal{S} = \mathcal{S}(\Phi,t, \xi, \eta),
\]
in other words, the instantaneous dependency of $\mathcal{I}$ on $\phi$ has become a ``global'' dependency of $\mathcal{S}$ on $\Phi$. For the pulse,  we numerically evaluate \eqnref{eqn:Ppulse} with the scaled vector potential \eqnref{eqn:apulse2}. To perform the calculation, we use the Bakhvalov-Vasil'eva method \cite{bakhvalov68}, which we briefly detail in \appref{App:A}.

\begin{figure}[h!!]
\includegraphics[width=8cm]{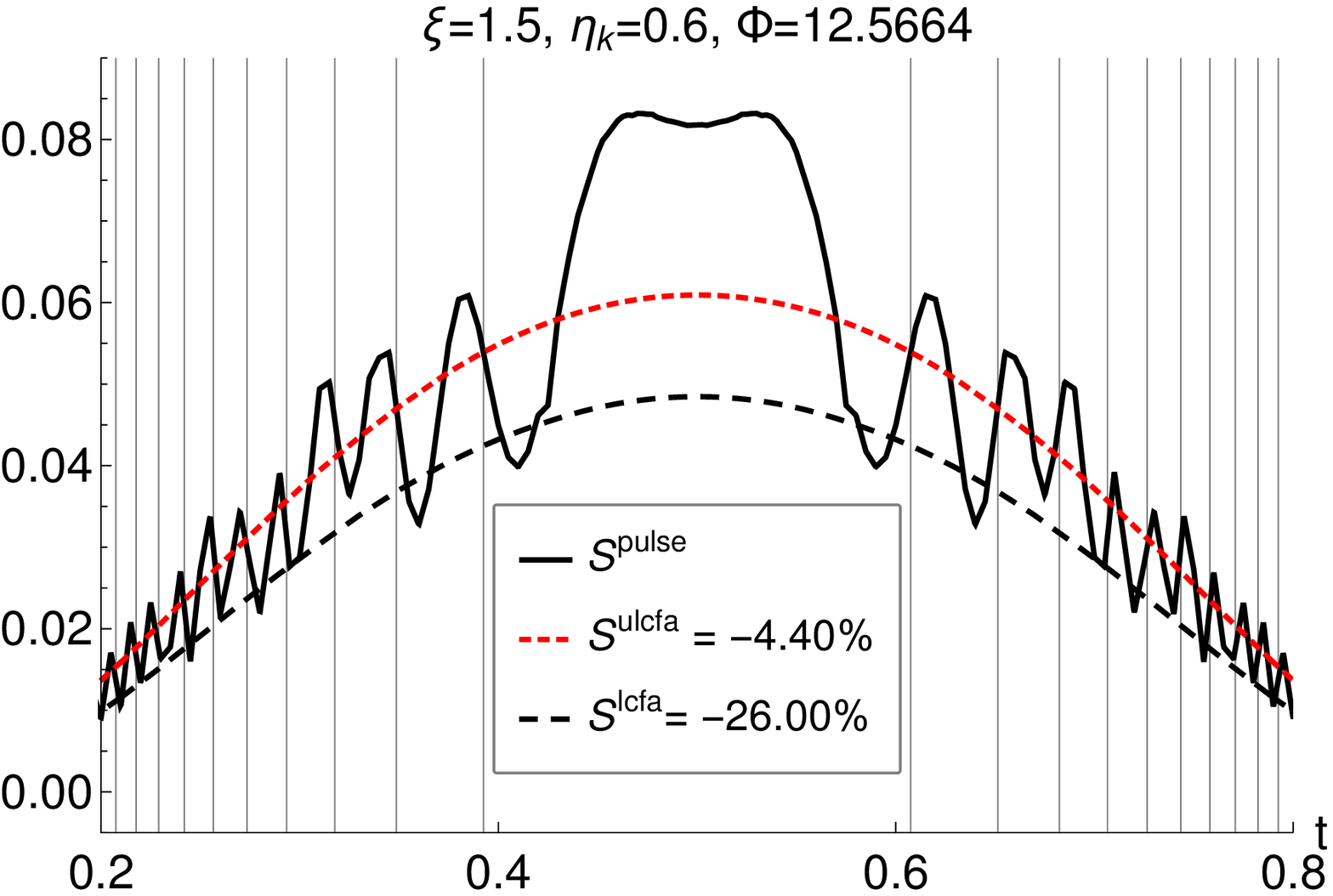}\\[3ex]
\includegraphics[width=8cm]{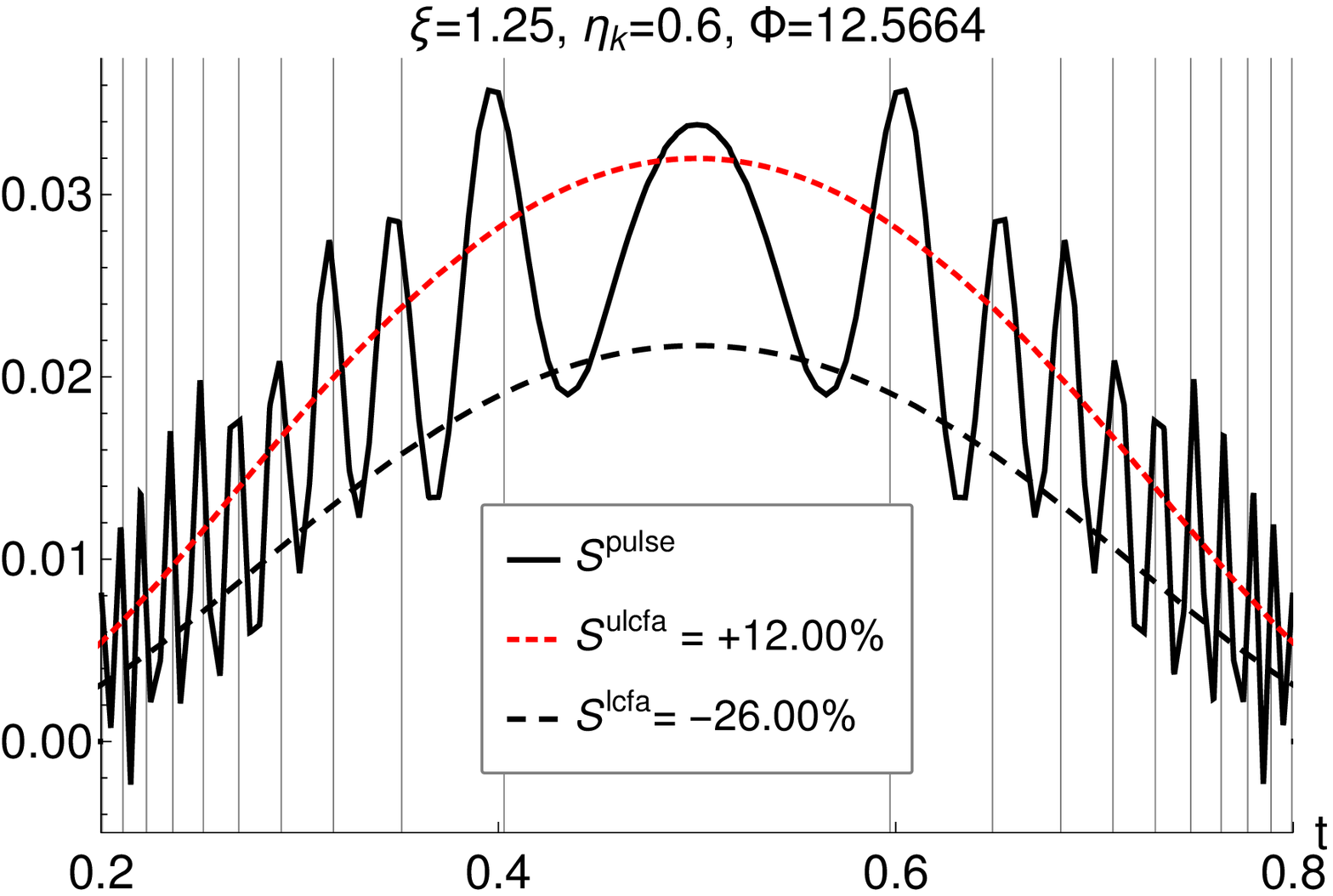}\\[3ex]
\includegraphics[width=8cm]{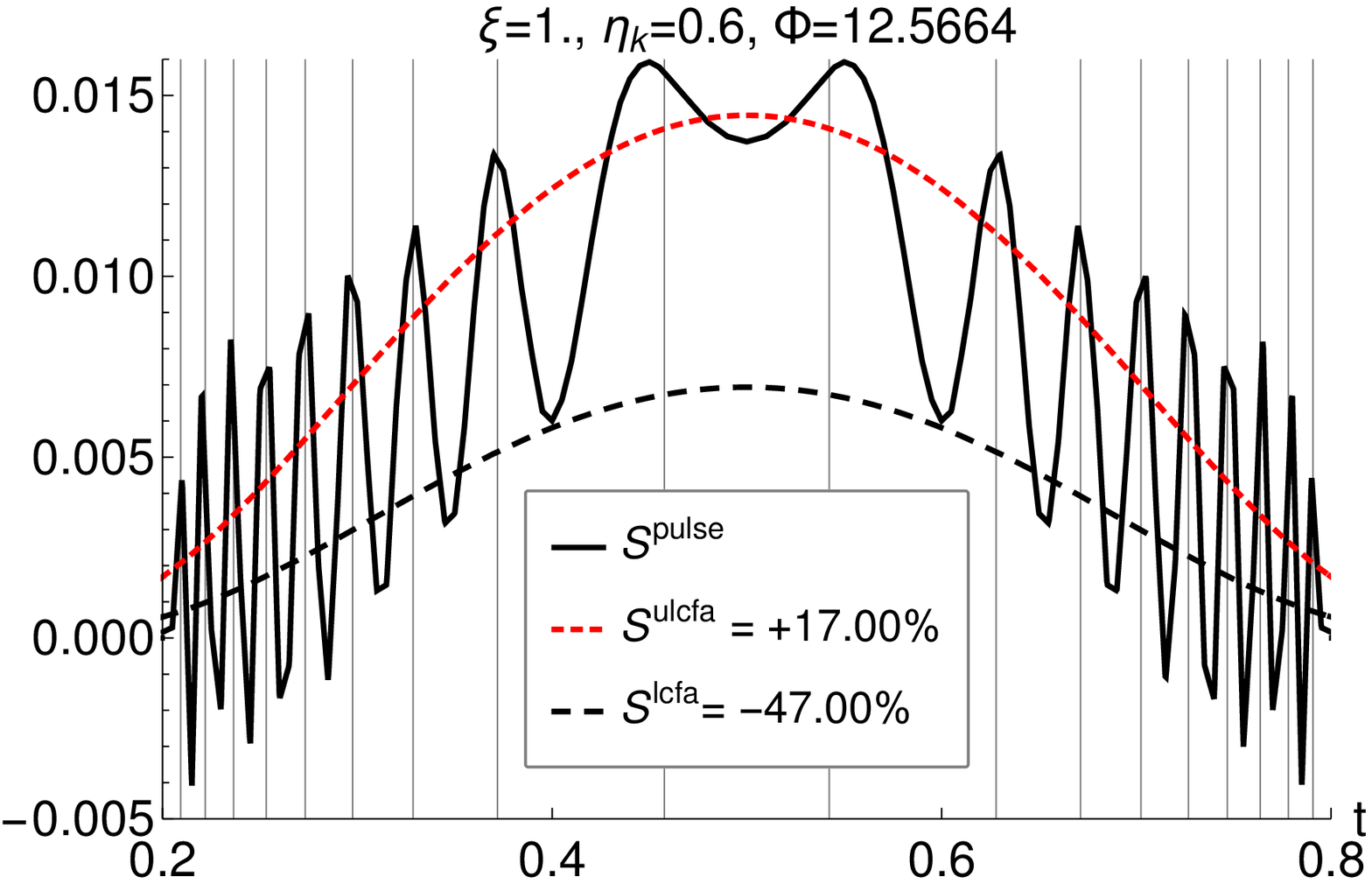}\\[3ex]
  \caption{Example plots of spectra in a pulse, showing the interference structure ($\eta_{k}=0.6$ corresponds to a frequency-tripled optical background). The harmonic subpeaks are correctly predicted by taking a cycle-average of the momentum relation \eqnref{eqn:C1} for integer multiples of the fast laser pulse frequency. } \label{fig:pulseres1}
\end{figure}

Some example spectra are plotted in \figref{fig:pulseres1}, where we see that for finite pulses, the ULCFA also performs better than the LCFA. We note the general trend that, for higher intensity there is better agreement between the local approximations and the finite pulse result. We also note that we are able to successfully predict the position of the pulse resonance subpeaks by a simple kinematic analysis. Consider energy-momentum conservation in a quasi-periodic field:
\[
k+ C\vkap = P+Q,
\]
where $P$ and $Q$ are the quasimomenta of the electron and positron respectively $P = p + \vkap (\xi^{2}/2\vkap \cdot p)$ and likewise for $Q$, where $C$ is an initially unknown real number. Squaring both sides of the energy-momentum conservation equation and assuming that, on average, $\mbf{p}^{\perp} = \mbf{q}^{\perp} = \pmb{0}$, one acquires an equation to solve for the lightfront momentum fraction $t$:
\bea
C \eta_{k} = (1+ \xi^{2})\left[ 1 + \frac{1-t}{2t} + \frac{t}{2(1-t)}\right]. \label{eqn:C1}
\eea
Now, this is very similar to the case of a circularly-polarised monochromatic background and making the comparison turns out to be useful. The main differences are, that for a finite pulse, $C$ is not an integer, and $\xi^{2}$ is not a constant. Therefore, it can be the case that the harmonic structure in a finite pulse is not well defined. This is clear, because to acquire a perfectly harmonic structure in the monochromatic derivation, one assumes that the background is constant in amplitude (and hence infinite in extent). This enters in the calculational step:
\[
 \sum_{n,n'}F_{n,n'} \lim_{L\to\infty} \int_{-L/2}^{L/2}\, d\phi~\mbox{e}^{i\phi (n-n')} \to 2\pi \sum_{n} F_{n,n},
\]
which is an integration over the external-field phase $\phi$, where $n$ and $n'$ are integers arising from decomposing the monochromatic background using the Jacobi-Anger relation \cite{landau4} and $F_{n,n'}$ is the phase-independent probability rate. This turns a double harmonic sum, which includes interference between  harmonics, into a purely harmonic expansion. Should either of the assumptions be false, the periodic symmetry of the background would be broken, and we should expect any harmonics to be ``smoothed out'' by interference. In \eqnref{eqn:C1}, we can take the cycle average of $(1+\xi^{2})$ over the ``fast'' part of the oscillation:
\[
 1 + \xi^{2}(\vphi) \to \frac{1}{N\pi}\int_{-2\pi N}^{2\pi N} 1 + \xi^{2}(x)~ dx = 4\left(1 + \frac{3\,\xi^{2}}{16}\right),
\]
and if we take the constant $C$ to be an integer, i.e. the momentum supplied by the laser field to be an integer multiple of the ``fast'' oscillation momentum, then we find the agreement given by the vertical grey lines in  \figref{fig:pulseres1}. However, we note that this is simply a useful observation, not a universal rule - when longer or more intense pulses were used, the agreement was not as clear as implied here, but the agreement demonstrates that the numerical routine is behaving in a logical way.
\newline

%
%
%
%
%
%
%
%
%
%

\section{Conclusion}
We presented a uniform locally constant field approximation (LCFA) for nonlinear Breit-Wheeler pair-creation in a plane wave background.  We displayed results where, for $\xi \sim O(1)$, our ``ULCFA'', was an order of magnitude more accurate than the LCFA in modelling the lightfront spectrum of pairs generated in the cases investigated: i) a circularly-monochromatic background and ii) a finite short pulse. The reason for the better agreement is that the ULCFA includes higher order derivatives of the background and therefore samples a larger region of the field than the standard LCFA. This extension is in the spirit of \cite{king19a}, but unlike \cite{king19a}, a uniform approach was taken and all corrections are within the Airy function arguments of the LCFA. This has the advantage that only one filter needs to be applied, namely an intensity filter which ensures that, when the intensity is weak enough, the LCFA is chosen over the ULCFA. For the short pulse spectrum, we demonstrated a kinematic argument that successfully predicted the positions of harmonic resonant subpeaks due to the finite pulse structure. 

The motivation for our work stems from upcoming experiments such as LUXE \cite{altarelli19} at DESY and E320 \cite{slacref1} at FACET-II using high-quality accelerator electron beams of the order of $10\,\trm{GeV}$, which will be collided with laser pulses with a moderate intensity parameter $\xi \sim O(1)$ to study high-intensity QED. We have demonstrated that LCFA methods can be extended to lower intensities than the standard applicability regime $\xi \gg 1$ and showed that the ULCFA can be accurate to within $10\%$ even down to intensities as low as $\xi \approx 1.25$.

\section*{Acknowledgments}
BK thanks Anton Ilderton for a careful reading of the manuscript. BK was partially funded from Grant No. EP/S010319/1.

\appendix

\section{Numerical Evaluation of Finite Pulses} \label{App:A}
We wish to evaluate integrals of the following type:
\bea
\mathcal{I} = \int_{-\theta_{f}}^{\theta_{f}}\,d\theta~f(\theta)\,\e^{icg(\theta)},
\eea
where $g(\theta)$ is monotonic in $\theta$ and antisymmetric: $g(-\theta) = - g(\theta)$ and $c$ is some positive non-zero constant. We begin by linearising the exponent, with the substitution: $g(\theta) = \gamma y$, where $\gamma=g(\theta_{f})$ to bring the integral in the form:
\bea
\mathcal{I} = \int_{-1}^{1}\,dy~H(y)\,\e^{ic\gamma y}; \quad H(y) = \frac{\gamma\,f[g^{-1}(\gamma y)]}{g'\left[g^{-1}(\gamma y)\right]}.\nn \\
\eea
At this point, we wish to use the method from Bakhvalov and Vasil'eva \cite{bakhvalov68} which we were directed to from \cite{olver08}, which utilises the result:
\[
 \int_{-1}^{1}dx\,P_{k}(x)\,\e^{i\omega x} = i^{k} \left(\frac{2\pi}{\omega}\right)^{1/2}J_{k+\frac{1}{2}}(\omega),
\]
for Legendre polynomials of degree $k$, $P_{k}(x)$, and hence decompose $H(y)$ using the standard result \cite{arfken12}:
\[
 H(y) = \sum_{k=0}^{k_{\tsf{max}}} \frac{2k+1}{2} P_{k}(y) \int_{-1}^{1} dy\, H(y)\,P_{k}(y),
\]
where $k_{\tsf{max}}$ is determined by a convergence criterion in the numerical evaluation. 
\newline

There is one subtelty with applying this method to QED processes in a pulse such as given in \eqnref{eqn:Ppulse}. If one na\"ively applies this method, it will fail due to the divergent $1/\theta$ pre-exponent term. However, the ``$i\eps$'' prescription gives a factor $-\pi/2$ from the $\theta$ integral, which we can rewrite and combine with the na\"ively-divergent term in the following way (see also \cite{dinu14a}). 
\bea
\mathfrak{I} &=& \int_{-\infty}^{\infty}\,d\theta~\frac{1}{\theta}\,\e^{icg(\theta)} - \frac{\pi}{2} \nn \\
&=& \int_{-\infty}^{\infty}\,d\theta~\frac{1}{\theta}\,\e^{icg(\theta)} - \int_{0}^{\infty}\,d\left[g(\theta)\right]\,\frac{\sin cg(\theta)}{g(\theta)}\nn \\
&=& \int_{-\infty}^{\infty}\,d\theta~\left[\frac{1}{\theta}-\frac{g'(\theta)}{g(\theta)}\right]\,\e^{icg(\theta)}\nn \\
&=& \int_{-1}^{1}\,dy~\left[\frac{1}{g'[g^{-1}(\gamma y)]}\frac{1}{g^{-1}(\gamma y)}-\frac{1}{y}\right]\,\e^{ic\gamma y},\nn \\
\eea
(the change in integral limits is made in recognition of the integrand disappearing outside of the pulse). At this point, the Bakhvalov Vasil'eva method can then be applied. Examples of our results using this method, are given in the main body of the paper in \figref{fig:pulseres1}

\bibliography{current}

\end{document}